\documentclass{article}
\usepackage{epsf}
\usepackage{emulateapj}
\usepackage{flushrt}
\usepackage{graphicx}

\def\etal    {{et~al.}~}

\def\arcmin{$^\prime$}

\lefthead{Donnelly, \etal}
\righthead{A HOT SPOT IN COMA}

\slugcomment{accepted by {\em The Astrophysical Journal}}
\begin{document}

\title{A Hot Spot in Coma}

\author{R.~Hank~Donnelly\footnotemark[1],
M.~Markevitch\footnotemark[1]$^{,2}$, W.~Forman\footnotemark[1],
C.~Jones\footnotemark[1], E.~Churazov\footnotemark[2]$^{,3}$ and
M.~Gilfanov\footnotemark[2]$^{,3}$}

\footnotetext[1]{Harvard-Smithsonian Center for Astrophysics,
Cambridge, MA 02138} 
\footnotetext[2]{Space Research Institute (IKI), Moscow 117810,
Russia} 
\footnotetext[3]{Max Planck Institute f\"{u}r Astrophysik, 85740
Garching bei M\"{u}nchen, Germany}


\centerline{\date}
\begin{abstract}
We study the temperature structure of the central part ($r\leq
18$\arcmin $\simeq 0.7 h_{50}^{-1}$ Mpc) of the Coma cluster of
galaxies using {\em ASCA} data. Two different analysis methods produce
results in good agreement with each other and reveal the presence of
interesting structures in the gas temperature distribution. Globally,
the average temperature in the center of the cluster is $9.0 \pm 0.6$
keV in good agreement with previous results. Superimposed on this, we
find a cool area with temperatures of 4-6 keV associated with a
filament of X-ray emission extending southeast from the cluster center
detected by Vikhlinin and coworkers. We also find a hot spot with a
temperature of around 13 keV displaced north from the central peak of
emission. The distribution of the gas temperatures and relative
specific entropies suggests that the cool features are most likely gas
stripped from a galaxy group centered on NGC 4874 falling toward the
core from outside, while the hot spot located ``ahead'' of this
in-falling gas is due to shock heating. Thus our results suggest that
we are observing Coma during a minor merger with a small group of
galaxies associated with NGC 4874 shortly before the initial core
passage.

\end{abstract}


\section{INTRODUCTION} 
The rich, nearby Coma cluster (A1656) has long been considered as the
archetype of a relaxed cluster system. However, with the advent of
X-ray imaging satellites ({\em Einstein}\/ and {\em ROSAT}\/), as well
as extensive optical studies, it has become clear that Coma, like many
other clusters (e.g., Geller \& Beers 1982; Jones \& Forman 1983; Bird
1994) exhibits significant substructure at various linear scales.

At the largest scales ($\sim 3\ h_{50}^{-1}$ Mpc), {\em ROSAT}
observations have shown a large subcluster $\sim 50$\arcmin\ to the
southwest of the center associated with NGC 4839 (Briel \etal 1992;
White \etal 1993; Burns et al.\ 1994). Optical studies ( Mellier \etal
1988, Biviano \etal 1996, hereafter B96; Colless \& Dunn 1996,
hereafter CD96) have found significant enhancements in the density of
galaxies associated with this subcluster. At intermediate($\sim 400
h^{-1}_{50}$ kpc) and smaller scales these authors as well as several
more (Fichett \& Webster 1987; White \etal 1993; Mohr, Fabricant, \&
Geller 1993) have found significant substructure in the X-rays and
complex velocity distributions in the optical and associated with the
two dominant galaxies NGC 4874 and NGC 4889.

Several authors (Davis \& Mushotzky 1993; White \etal 1993; Vikhlinin
\etal 1994) have found distinctive enhancements of X-ray emission
associated with the galaxy concentrations around the two dominant
galaxies, NGC 4889 and NGC 4874, for which they obtained mass
estimates consistent with those from the velocity dispersion inside
these galaxy groups (Mellier \etal 1988). A more detailed wavelet
analysis of the {\em ROSAT}\/ image (Vikhlinin \etal 1997, hereafter
V97) has also revealed a gas filament extending southeast from the
cluster core for $\sim 1\,h_{50}^{-1}$ Mpc. One explanation for this
filament is as a trail of gas left in the wake of an in-falling galaxy
group.

Temperature maps of the X-ray emitting gas are a new and especially
sensitive tool for the detection of dynamic activity. Hydrodynamic
simulations indicate that mergers should produce characteristic
temperature patterns that survive 4-6 times longer than perturbations
in the gas density (e.g.  
 Schindler \& M\"uller 1993; Ricker 1997). Recent observations
(Henry \& Briel 1995; Henriksen \& Markevitch 1996; Markevitch \etal
1996b, 1998; Donnelly
\etal 1998) have found just such characteristic temperature structures
in a variety of clusters, which have been interpreted as due to
cluster mergers. 

A large-scale ($r=60^{\prime}\simeq 2.4\,h_{50}^{-1}$ Mpc) temperature
map of Coma has been presented by Honda \etal (1996), who used GIS
data from fourteen different {\em ASCA}\/ pointings. Their map has a
resolution of $30'$ (a GIS field of view) and reveals a hot region
$40'$ northwest of the cluster as well as several other significant
deviations from isothermality. Watanabe \etal (1997) presented a
preliminary analysis of the temperature structure at smaller scales.

In this paper, we use the {\em ASCA} observation of the Coma center to
obtain a more detailed, $\sim 5$\arcmin\ resolution, temperature map
of the central ($r\leq 18$\arcmin $\simeq 0.7 h_{50}^{-1}$ Mpc) region
of the cluster.  All distance dependent quantities have assumed $\rm
H_o= 50\ km\ s^{-1}\ Mpc^{-1}$, $q_o=0.5$, all coordinates are given
in the J2000 system, and unless otherwise noted all error bars are
90\% confidence level.
\section{OBSERVATIONS \& RESULTS}
{\em ASCA} observed the core of the Coma cluster on June 14, 1993
(sequence \# 80016000), and had useful exposures of 7.6, 7.6, 9.4 and
9.4 ksec for the SIS 0-1 and GIS 2-3 detectors respectively. We have
not included the thirteen offset pointings centered on this cluster (see
Honda \etal 1996) because they are subject to ``stray light''
contamination by photons scattered from the central area which we do
not correct for in our analyses.

\begin{figure*}[t] 
\includegraphics[totalheight=3.00in]{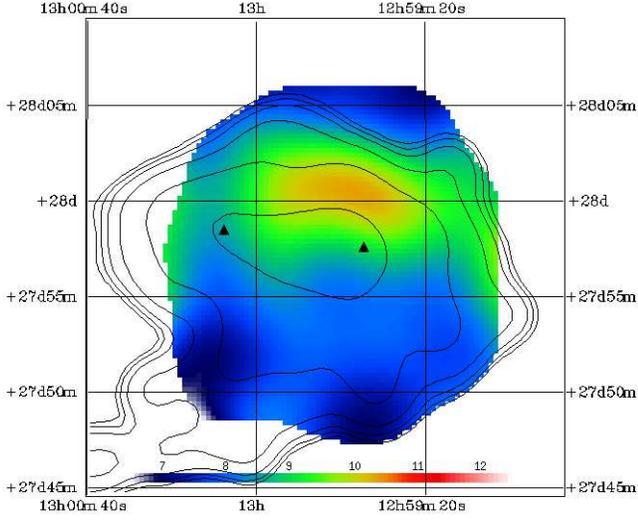}\includegraphics[totalheight=3.in]{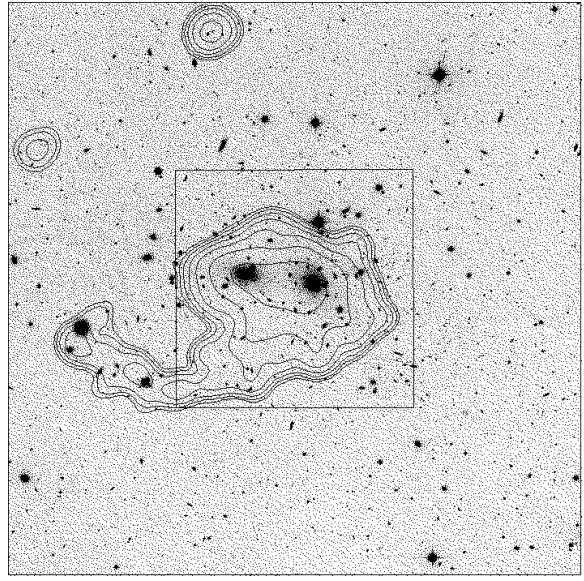}
\caption{Panel a shows the continuous temperature map produced using
Method A (the scale bar is in keV). The map excludes regions where the
uncertainty in the temperature is large, $T+\sigma_T/T-\sigma_T
\geq 1.5$).  Contours in both panels show the small scale structure in
the Coma {\em ROSAT} PSPC image, detected by V97 after the removal of
the main cluster emission using wavelet smoothing. In Panel b
(reproduced from V97), the contours are overlaid on the Digital Sky
Survey image. A square outlines the area from Panel a. The two
central galaxies, NGC 4874 and 4889, are on the right and left
respectively and have been marked as triangles in Panel a.}
\label{fig:e+mmap} 
\end{figure*}

Correctly characterizing the temperature distribution from {\em ASCA}
data for an extended source on scales smaller than the entire field of
one of the detectors requires a correction for the energy dependent
PSF of the telescope (Takahashi \etal 1995).  To this end, we have
employed two independent methods detailed elsewhere (Churazov \etal
1996, 1998, hereafter Method A; Markevitch \etal 1996a, 1998, hereafter
Method B) to generate temperature maps with resolutions of $\sim
5$\arcmin. Method A provides a continuous temperature map, although
the correction for the PSF is approximate (see Churazov \etal 1998 for
details). Method B derives spectra in projected regions defined by the
user and takes exact account of the {\em ASCA} PSF. Note that both
methods use the same {\em ASCA} PSF model and therefore may be subject
to the same systematic uncertainties.

Figure~\ref{fig:e+mmap}a, developed using Method A applied to the {\em
ASCA} GIS data, shows the approximate best fit temperature, excluding
the outer regions of the field of view where the uncertainty in the
temperature was large. The map has been smoothed to a spatial
resolution of $\sim 5$\arcmin\ to improve the signal-to-noise
ratio. Contours in both panels of Figure~\ref{fig:e+mmap} show the
filamentary structure in the X-ray intensity revealed by V97 after the
removal of large-scale cluster emission via wavelet smoothing of the
{\em ROSAT} PSPC image (see Vikhlinin \etal 1994 and V97 for details).
In Figure~\ref{fig:e+mmap}b the wavelet contours are overlaid on the
optical image from the Digital Sky Survey. We have added a box to
delineate the area shown in Figure~\ref{fig:e+mmap}a. The positions of
the two especially prominent galaxies, NGC 4874 and NGC 4889, are
indicated in Figure~\ref{fig:e+mmap}a with solid triangles. There are
three prominent features in the temperature map: two areas $\sim 20$\%
cooler than the average on the south and southeast edges of the map
coincident with the filament, and a hot spot, $\sim 25$\% hotter,
displaced $\sim 5$\arcmin\ to the north from the peak of the intensity
located $\sim 1.5$\arcmin\ southeast of NGC 4874.

\begin{figure*}[t]
\includegraphics[totalheight=3.3in]{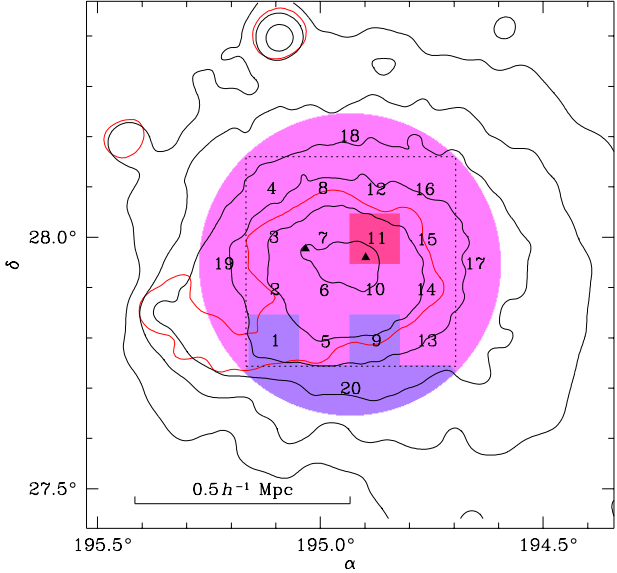}\includegraphics[totalheight=3.9in]{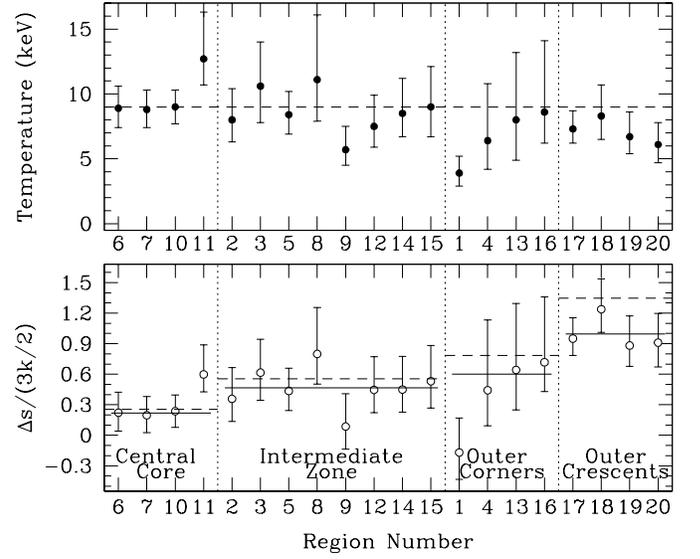}
\caption{The binned temperature map produced by the Method B. The
black contours show the intensity from the {\em ROSAT} PSPC data and
the red contour is the lowest contour shown in
Figure~\ref{fig:e+mmap}.  Different colors indicate significant
differences in temperature. The temperatures for each region with 90\%
uncertainties are plotted in the upper right panel. The lower right
panel shows the estimated relative specific entropies for each region
(see text for details). In the temperature panel, the average cluster
temperature is shown as a dashed line. Dashed lines in the specific
entropy panel show the entropy values for a symmetric, isothermal
$\beta$-model with $T=9$ keV, relative to its central value, while the
solid lines indicate average observed values in the groups, excluding
regions 1 and 11 which are obviously discrepant. The positions of NGC
4874 and NGC 4889, from right to left in the cluster core, are shown
as filled triangles on the map, and a box has been provided to show
the area displayed in Figure~\ref{fig:e+mmap}a.}
\label{fig:maximmap} 
\end{figure*}

To better understand the statistical significance of the thermal
deviations found in Figure~\ref{fig:e+mmap}, we employed Method B to
produce a binned map of the temperature (Figure~\ref{fig:maximmap})
using data from both the GIS and SIS detectors. Integral to this
method is the use of an intensity model which in this case we have
chosen to be the {\em ROSAT} PSPC image. A 30\arcsec\ relative offset
between the {\em ROSAT} and {\em ASCA} images was corrected so that
model and data would be correctly aligned. The lowest wavelet contour
from Figure~\ref{fig:e+mmap} has been overlaid in the figure in red,
the {\em ROSAT} PSPC contours are overlaid in black, and the positions
of NGC 4874 and NGC 4889 are marked with filled triangles. The area
displayed in Figure~\ref{fig:e+mmap}a is outlined for comparison. A
plot of the temperature versus region is provided in the upper right
panel of Figure~\ref{fig:maximmap}.

The integrated spectrum from within a 9\arcmin\ radius of the cluster
center yields a best fit average temperature of $9.0\pm 0.6$ keV,
consistent with previous results (Hughes \etal 1993, Honda \etal
1996). Figure~\ref{fig:maximmap} shows that, except for regions 1, 9,
11 and 20, there are no significant deviations from this
average. Region 11 ($\rm kT=12.7^{+3.6}_{-2.0}$ keV) coincides with
the center of the hot spot found in Figure~\ref{fig:e+mmap}, while
regions 1, 9 and 20 ($\rm kT= 3.9^{+1.3}_{-1.0},\ 5.7^{+1.8}_{-1.2}\
and\ 6.1^{+1.7}_{-1.4}$ respectively) generally coincide with the cool
regions along the filament in Figure~\ref{fig:e+mmap}. As with
previous work (Donnelly \etal 1998), the temperatures determined by
both methods are in good agreement.

\section{DISCUSSION} 
\label{sec:result}
The simplest interpretation of our temperature maps, in agreement with
B96, is that Coma is experiencing a merger with a moderately sized
group associated with NGC 4874 located south of the observed hot
spot. The hot spot of emission is due to a bow shock generated by the
group as it moves through Coma's intercluster medium(ICM). As the
group has fallen toward the core of Coma, it has left a trail of cool
gas delineating its in-bound track, which we see as the filament
detected by V97. This is similar, for example, to simulations by
Roettiger, Loken \& Burns (1997, hereafter RLB97), Roettiger, Burns \&
Loken (1996, hereafter RBL96) and Norman \& Bryan (1998, hereafter
NB98).

Recently both CD96 and B96 reported significant evidence for
substructure in the core of Coma associated with the galaxies
surrounding both NGC 4889 and NGC 4874. CD96 argued that the group
around NGC 4874 was the ``true central peak'', and the outer halo of
the cD-- which would have been destroyed during core passage- has been
re-aquired from merger debris during its long term residence in the
core. They were, however, hard pressed to explain the large velocity
offset ($\sim 350$ km/s) relative to the overall field required from
their model for NGC 4874.

Alternatively, the presence of a cD galaxy in the form of NGC 4874
could indicate that the group is in an early stage of the merger prior
to entering the core, and as such has not yet been disrupted by the
merger process.  B96 separated the cluster galaxies into ``bright''
and ``faint'' samples, and found-- within the ``bright'' sample-- two
galaxy groups, one associated with each of the two dominant
galaxies. This is similar to the results of Fitchett \& Webster
(1987).  B96 then argued that the ``faint'' sample was a better
representative of Coma's overall velocity field and as such indicated
that the two groups were ``still in the process of merging with the
cluster''. This model would explain the presence of the cD, the
velocity offset for the NGC 4874 group as well as the temperature
structures that we have found and the offset of the peak of Coma's
overall emission from the two dominant galaxies (White \etal 1993,
Vikhlinin \etal 1994).

Following Markevitch \etal (1996b), we used the temperature maps to
estimate the local relative specific entropy (entropy per particle) to
better understand the gas merger process and the energetics
involved. The relative specific entropy is defined as
\begin{equation}
\Delta
s=s-s_0=\frac{3}{2}k\ln\left[\frac{T}{T_0}\left(\frac{\rho}{\rho_0}\right)^{-\frac{2}{3}}\right],
\label{eq:entropy}
\end{equation}
where $\rho_0$, $T_0$ and $s_0$ are, in this case, the central values
from an isothermal $\beta$-model ($\rho_{model}\propto
\rho_0(1+r^2/r_c^2)^{-3\beta/2}$, Cavaliere \& Fusco-Femiano
1976).  The emission-weighted average density in each projected region
from the symmetric $\beta$-model was crudely corrected for the
cluster's observed asymmetry by multiplying by the square root of the
ratio of the actual brightness, as determined from the {\em ROSAT}
data, to the predicted model brightness:
\begin{equation}
\rho=\rho_{model}\times\left(\frac{\Sigma_{obs}}{\Sigma_{model}}\right)^{\frac{1}{2}},
\label{eq:sbcor}
\end{equation}
where $\rho_{model}$ is the value from the $\beta$-model.  This allows
us to compare the specific entropy for regions of interest via the
common benchmark of the $\beta$-model, while taking into account the
smaller scale non-uniformities. We used the best fit $\beta$-model
parameters $\beta=0.75$ and core radius $r_c=10.5$\arcmin\ from Briel
\etal (1992).

The lower right panel in Figure~\ref{fig:maximmap} shows the relative
specific entropy in each region with its 90\% confidence error
bar. The errors include contributions from the uncertainties in the
surface brightness correction and the $\beta$-model, but are strongly
dominated by the contribution from the temperature determination. For
clarity, we have segregated the regions into four groups, such that
the relative specific entropy for the azimuthally symmetric isothermal
model (shown as dashed lines) is constant within each group. The
average relative specific entropy for each group (solid lines) is
shown for comparison. Due to the obviously discrepant nature of both
regions 1 and 11, they were excluded from the average for their
respective groups.

In the core, Region 11 (the hot spot) has a significantly higher
relative specific entropy compared to the other central
regions. Region 8 also has a high relative specific entropy relative
to its group, although this deviation is only marginally
significant. Assuming that this hot gas lies inside the core, and that
initially it had a relative specific entropy equal to that of the rest
of the core, we can estimate the total energy that has been injected
into the gas as:
\begin{equation} Q\simeq T\Delta S \sim 0.3-1.0 \times 10^{61}
h^{-\frac{5}{2}}_{50} ergs.
\end{equation}
By comparison, a gas mass of $2\times 10^{12}\ M_\odot$ associated
with a group falling from 5 Mpc into the core will acquire
approximately 10-30 times this amount in kinetic energy, part of which
it should lose prior to reaching the core. The conversion of the gas's
gravitational potential energy is then a reasonable source for the
energy required to produce the hot spot.

We find that the temperatures in the cool regions (1, 9 and 20) are
consistent with the allowed lower limit of 4 keV calculated by V97 for
the filament. Regions 1 and (less significantly) 9 both also have low
relative specific entropies compared to their group averages.  Region
1 is especially notable in that it has a much lower relative specific
entropy than even the core values of the $\beta$-model. The low values
of the relative specific entropy for regions 1 and 9 suggest that the
gas in these regions only recently began to fall into the cluster and
has not yet been completely mixed with or heated by the cluster's
ICM. These regions are all spatially associated with the filament seen
in X-ray image.

In their analysis of the filamentary emission, V97 offered two
possible explanations for the filamentary enhancement in
emission. Either the filament arises from gas stripped via ram
pressure from a group passing through Coma, or from a
local perturbation in the ICM due to dark matter that has been tidally
stripped from a merging group. The distinctive relative specific
entropy of gas in the filament supports the former explanation.

RLB97's simulations of a merger with a small group (see also RBL96 and
NB98), find that the passage of a group through the cluster environment will
lead to substantial vorticity along the path of the in-bound
object. This would temporarily introduce an additional pressure term
in the in-falling gas preventing it from thermally equilibrating with
the overall ICM. It is also possible that the vorticity would retard
the mixing of the in-falling gas into the overall ICM, thus prolonging
the lifetime of the filament and addressing some the time scale issues
raised in V97.

On smaller scales, numerical simulations (RBL96, RLB97) indicate that,
in a minor merger, the in-falling object will develop a protective bow
shock, which will effectively shelter the {\em galaxies} from being
ram pressure stripped of their gas. However, close to core passage the
in-falling object's ICM encounters a rapidly increasing density in the
gas of the main cluster. This would lead to the group's gas and
galaxies decoupling from each other, and the galaxies experiencing an
``impulsive burst of ram pressure'' as they pass through the
shock. This would naturally lead to the galaxies being stripped of
their gas prior to their coalescence with the cluster. Contrary to
this, Vikhlinin \etal (1994) find a large gas halo associated with NGC
4874, while Schombert (1988) finds the optical completely normal and
unperturbed for a cD galaxy.

Radio observations of the galaxy NGC 4869, $\sim 4^{\prime}$ southwest
of NGC 4874, by Feretti \etal (1990) indicate that it is in orbit
around the cD galaxy. They noted a major problem in the confinement of
the radio tail associated with NGC 4869 due to its internal pressure
strongly decreasing with distance from the center of the galaxy.  One
possible explanation is that, due to a projection effect, the pressure
of the ambient medium is lower than that which they have computed.
Feretti \etal (1990) reject this based upon the assumption that NGC
4874 is nearly at rest at the bottom of the cluster potential. If,
however, NGC 4874 and NGC 4869 are part of a group still falling into
Coma a significant projection effect may be at work.  Given the large
extent of the X-ray gas-- more than 12\arcmin\ in diameter-- found by
V97 around NGC 4874, it is plausible that NGC 4869 is being shielded
from the overall cluster ICM (RLB97), and the relevant ambient medium
is that of the in-falling group.

These arguments suggest that the group has not quite arrived at the
core. Further, the velocity studies of Mellier \etal, B96 and CD96
find a statistically significant offset for the NGC 4874 group
relative to the Coma's average velocity field. This suggests that that
the group is falling in from the front side of Coma at a mild
inclination to our line of sight.

There are some complications to this simple model due to the nearby
presence of another subcluster associated with NGC 4889, $\sim
7$\arcmin\ to the east from NGC 4874. Baier \etal (1990) find that the
group of galaxies associated with NGC 4889 is elongated along an axis
from the southeast to the northwest, possibly indicating a recent
dynamical/tidal interaction similar to that described by RLB97 as
occurring after core passage. It is possible that the slightly high
entropy in region 8 briefly mentioned above is related to this
activity. The lack of an extended envelope around NGC 4889 also would
be expected if the group has recently passed through the core.  As
with the NGC 4874 group, CD96 and B96 find a statistically significant
velocity offset for the NGC 4889. This suggests that {\em both} the
NGC 4889 and NGC 4874 groups -- in that order-- may be recent arrivals
in the core. We are thus seeing the debris of the final stages of one
minor merger, just as another minor merger is occurring.

Other complications arise from what the model described above suggests
that we {\em should} see. Nominally, we expect that the surface
distribution of galaxies in an in-falling group would experience some
distortion prior to core passage. In their simulations, RLB97 show
just such a morphology at -0.25 Gyr for a 2:1 mass ratio merger,
although the distortion is less for the secondary (i.e. in-falling)
cluster than for the primary and it is not clear how this effect would
scale with mass.  However, Baier \etal (1990) find that the group of
galaxies around NGC 4874 is circularly distributed suggesting no
dynamical perturbations.  This lack of distortion evident in the NGC
4874 group may be due to its being still quite far from core passage.

Finally, we would expect the temperature and entropy results for
Region 5-- situated directly between regions 1, 9 and 20-- to share
in the properties of the filament as outlined above. However, we find
that it is nearly identical to the non-filament gas. Perhaps this is
due to a projection or dilution effect due to the increasing
brightness of the overall ICM. Unfortunately, the resolution of our
data is inadequate for further speculation.

\section{SUMMARY}
We have analyzed the {\em ASCA} pointing covering the central
r=18\arcmin\ area of the Coma cluster. Although, the temperature map
of the gas in the core of the cluster does not exhibit the strong
temperature variations observed in clusters undergoing major mergers,
we find several significant features in the temperature map indicating
recent and ongoing mergers with small groups.

Our analysis indicates that gas to the south/southeast side of the
cluster is significantly cooler and has a lower relative specific
entropy than the average. The regions of low temperature and relative
specific entropy coincide with the gas filament, detected by V97,
extending from the cluster's center to the south and east towards 
NGC 4911. This low entropy gas is most likely deposited in the wake of the
NGC 4874 group as it falls toward the cluster.

The temperature map also reveals a hot spot located to the north of
the cD galaxy, NGC 4874, and the peak of the intensity profile. This
is most likely a bow shock in the gas which has developed as this
small group of galaxies has fallen in to the core of the cluster. Our
rough estimate of the relative specific entropy of the shocked gas is
consistent with heating from the conversion of the kinetic energy of a
small group falling into the cluster.

The X-ray and optical data suggest that a small group of galaxies
associated with NGC 4874 is in-bound towards the cluster core from the
southeast leaving a tail of cooler gas along its trajectory. The
relatively unperturbed state of the group suggests that core passage
has not yet occurred. The group associated with NGC 4889, although
still in the process of being incorporated into the cluster, has
already passed the core and most likely been resident in the core for
some time, as evidenced by its dynamically disturbed state.

Future observations of Coma with {\em AXAF} will allow us to study the
characteristics of the ICM around these merger remnants at much higher
spatial resolution. This will help us to understand the processes at
work during and after mergers in much greater detail.

\acknowledgments 
We would like to thank the referee for their helpful suggestions and
comments.  RHD, MM, WF, CJ acknowledge support from the Smithsonian
Institute and NASA contract NAS8-39073.

\end{document}